\documentclass[aps,showpacs,prl,twocolumn,superscriptaddress,monochrome]{revtex4-2}
\usepackage{graphicx}

\newcommand\Oone{O$_\mathrm{I}$ }
\newcommand\Otwo{O$_\mathrm{II}$ }

\usepackage{color}
\usepackage{siunitx}
\usepackage{chemformula}
\usepackage{dcolumn}
\graphicspath{{images/}}
\usepackage{hyperref}

\makeatletter
\def\maketitle{
\@author@finish
\title@column\titleblock@produce
\suppressfloats[t]}
\makeatother
\begin{document}

\title{
Unravelling the atomic and electronic structure of nanocrystals on superconducting Nb(110): Impact of the oxygen monolayer
}

\author{Samuel Berman}
\email[]{bermans@tcd.ie}
\affiliation{School of Physics and Centre for Research on Adaptive Nanostructures and Nanodevices (CRANN), Trinity College Dublin, The University of Dublin, Dublin 2, Ireland}
\author{Ainur Zhussupbekova}
\affiliation{School of Physics and Centre for Research on Adaptive Nanostructures and Nanodevices (CRANN), Trinity College Dublin, The University of Dublin, Dublin 2, Ireland}
\affiliation{School of Chemistry, Trinity College Dublin, The University of Dublin, Dublin 2, Ireland}
\author{Brian Walls}
\affiliation{School of Physics and Centre for Research on Adaptive Nanostructures and Nanodevices (CRANN), Trinity College Dublin, The University of Dublin, Dublin 2, Ireland}
\author{Killian Walshe}
\affiliation{School of Physics and Centre for Research on Adaptive Nanostructures and Nanodevices (CRANN), Trinity College Dublin, The University of Dublin, Dublin 2, Ireland}
\author{Sergei I. Bozhko}
\affiliation{Institute of Solid State Physics, Russian Academy of Sciences, Chernogolovka, Russia}
\author{Andrei Ionov}
\affiliation{Institute of Solid State Physics, Russian Academy of Sciences, Chernogolovka, Russia}
\author{David D. O'Regan}
\affiliation{School of Physics and Centre for Research on Adaptive Nanostructures and Nanodevices (CRANN), Trinity College Dublin, The University of Dublin, Dublin 2, Ireland}
\author{Igor V. Shvets}
\affiliation{School of Physics and Centre for Research on Adaptive Nanostructures and Nanodevices (CRANN), Trinity College Dublin, The University of Dublin, Dublin 2, Ireland}
\author{Kuanysh Zhussupbekov}
\email[]{zhussupk@tcd.ie}
\affiliation{School of Physics and Centre for Research on Adaptive Nanostructures and Nanodevices (CRANN), Trinity College Dublin, The University of Dublin, Dublin 2, Ireland}
\affiliation{School of Chemistry, Trinity College Dublin, The University of Dublin, Dublin 2, Ireland}

\begin{abstract}

The Niobium surface is almost always covered by a native oxide layer which greatly influences the performance of superconducting devices. Here we investigate the highly stable Niobium oxide overlayer of Nb(110), which is characterised by its distinctive nanocrystal structure as observed by scanning tunnelling microscopy (STM). Our \textit{ab-initio} density functional theory (DFT) calculations show that a subtle reconstruction in the surface Niobium atoms gives rise to rows of 4-fold coordinated oxygen separated by regions of 3-fold coordinated oxygen. The 4-fold oxygen rows are determined to be the source of the nanocrystal pattern observed in STM, and the two chemical states of oxygen observed in core-level X-ray photoelectron spectroscopy (XPS) are ascribed to the 3-fold and 4-fold oxygens. Furthermore, we find excellent agreement between the DFT calculated electronic structure with scanning tunnelling spectroscopy and valence XPS measurements.

\end{abstract}


\maketitle

\section{I. Introduction}

Superconducting electronics are a cornerstone of modern science and technology, from superconducting radio-frequency cavities (SRFs) in particle accelerators\cite{Kneisel2015, 
Wenskat2020, Wenskat2022}, to qubits in quantum computers \cite{Leek2009, Oliver2005}. Niobium has emerged as one of the most attractive materials for superconducting electronics due to its high transition temperature and critical field \cite{Asada1969}, as well as its ease of fabrication \cite{Ciovati2006}. Niobium samples typically contain a large amount of interstitial oxygen and quickly form an oxide layer when exposed to atmosphere \cite{Delheusy2008, Surgers2001}. While the superconducting gap of Niobium is sensitive only to the presence of bulk interstitial oxygen \cite{Koch1974, Odobesko2020_2, Veit2019, Prischepa2021}, device properties such as coherence time in Josephson junctions \cite{Harrelson2021, Nathelie2021}, qubit relaxation time \cite{Premkumar2021}, and quality factor in SRF cavities \cite{Ciovati2006, Romanenko2017, Semione2021, Prudnikava2022, Trenikhina2015, Verjuaw2021} are greatly affected by the precise nature of the surface oxide layer. For SRF cavities, it has been shown that annealing the sample in ultra-high vacuum conditions can compensate for the ``Q-drop'' \cite{Ciovati2010}, i.e. the sharp drop of in quality factor (Q-factor) above 80 mT \cite{Palmieri1998qh, Ciovati2010}. The Q-drop is believed to be a surface effect, with proposed mechanisms including local Joule heating of inhomogeneities with high surface impedance \cite{Gurevich2012}, or the motion of magnetic vortices pinned at the surface \cite{Gurevich2008}. Understanding the precise structure of the surface oxides is therefore vital for improving device performance.

\begin{figure*}
    \includegraphics[width=2\columnwidth]{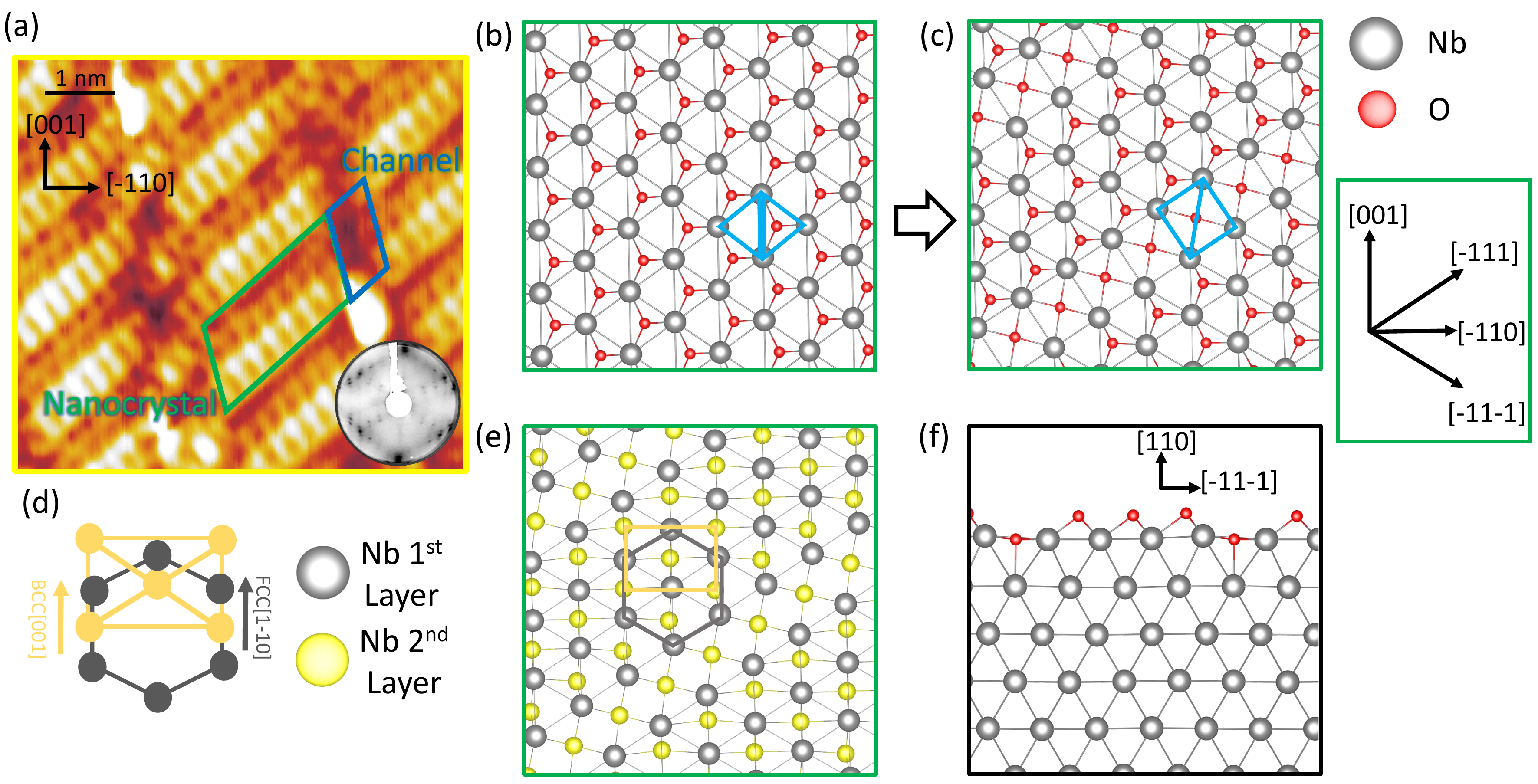}
    \caption{(a) STM image of the surface with $V$=2 mV, $I$=60 pA, showing the nanocrystal and channel regions of the surface (with measured LEED pattern inset). (b) Unrelaxed and (c) relaxed structures of a full monolayer of oxygen, showing the emergence of a 4-fold site (top view). (d) Schematic of the NW epitaxial relationship predicted from LEED experiments. (e) top view of 1st and 2nd Nb layers with oxygen removed to highlight the fcc-bcc epitaxial relationship. (f) Side view of relaxed geometry.
}

\label{fig:fig1}
\end{figure*}

The (110) surface of Niobium has been studied by a variety of experimental techniques, including scanning tunnelling microscopy/spectroscopy (STM/STS) \cite{Zhussupbekov2020, Razinkin2010, Arfaoui2002, Yazdani1997}, low-energy electron diffraction (LEED) \cite{Zhussupbekov2020, Hellwig2003}, grazing incidence X-ray diffraction (GIXRD) \cite{Delheusy2008,delheusy_2008}, Auger electron spectroscopy (AES) \cite{Arfaoui2002, Surgers2001}, and X-ray photoelectron spectroscopy (XPS) \cite{Hu1989, Razinkin2008, Zhussupbekov2020}. When exposed to atmosphere, a layered oxide structure of NbO, NbO$_2$, and amorphous Nb$_2$O$_5$ \cite{delheusy_2008} is formed. The NbO$_2$ and Nb$_2$O$_5$ layers can be removed by annealing in ultra-high vacuum (UHV), leaving only the NbO layer \cite{Delheusy2008}. The remaining oxide is very difficult to remove, requiring annealing with temperatures close to the melting point of Niobium, in excess of 2400 $^\mathrm{o}$C \cite{Odobesko2019, Odobesko2020_2, Boshuis2021, Beck2021, Beck2022}. This structure will of course greatly influence the electronic structure of the surface, as well as the structure of additional layers formed on top of the surface \cite{Zhussupbekov2020}. Our STM experiments on this surface show a distinctive pattern of finite rows of wide protrusions, consistent with previous reports \cite{Arfaoui2002, Razinkin2010, Zhussupbekov2020}. These rows of protrusions are termed the ``nanocrystals'', separated by disordered regions dubbed ``channels'' (see Figure \ref{fig:fig1}(a)).

Despite the technological and scientific relevance of the oxidised Nb(110) surface, very few calculations have been performed to understand its structure. While some previous works have performed calculations for the clean Nb(110) surface \cite{Lekka2003, Odobesko2019}, the oxidised surface has received sparse attention. Kilimis et al. \cite{Kilimis2007} performed DFT calculations studying the effect of 2-fold coordinated oxygen covering the surface, with a view towards understanding the early stages of the oxidation process. Other works have computed the minimum energy paths for diffusing oxygen atoms through the surface \cite{Wang2012,Tafen2013}. However, no previous reports have tackled the equilibrium structure or electronic properties of the nanocrystals, which plays a vital role in device performance. In this work, we investigate this surface structure by \textit{ab-initio} density functional theory (DFT) calculations, along with complementary STM, STS, LEED, XPS and UPS experiments. We show that this surface structure can be explained by a subtle oxygen induced reconstruction, with no need for Niobium adatoms. \textcolor{red}{In the supplemental material \cite{SI1} (see also references \cite{Efimenko2017,Hu2020,Schulz1993} therein) we show a similar analysis for models previously presented in the literature from Razinkin et al. \cite{Razinkin2010} and Arfouai et al. \cite{Arfaoui2004}, and demonstrate that these types of model cannot correctly explain the experimental data.}

\section{II. Methods}

\subsection{A. Computational Details}

DFT calculations were carried out utilising the Quantum Espresso plane-wave self-consistent field (PWscf) package \cite{Giannozzi2020, Giannozzi2009}. Throughout, the non-spin-polarized PBE exchange-correlation functional \cite{Perdew1996} and projector augmented wave (PAW) pseudopotentials \cite{DalCorso2014} were used. We utilise symmetric Nb(110) slabs with 9 atomic layers and vacuum padding of at least 10 \AA, along with a wavefunction cutoff energy of 45 Ry, \textit{k}-point sampling of 16$\times$16$\times$1 in the Nb[-111], Nb[-11-1], and Nb[110] directions respectively, and Marzari-Vanderbilt smearing \cite{Marzari1999} of 0.005 Ry giving a total energy convergence of $<$ 1 meV/atom. The number of \textit{k}-points was scaled when constructing the surface reconstruction according to the unit cell dimensions. Crystal structures were visualised in VESTA \cite{Momma2011}. Simulated STM images and STS curves were obtained within the Tersoff-Hamann approximation \cite{Tersoff1983, Tersoff1985}, by integrating the local density of states (LDOS) over a small 0.037 \AA$^3$ volume at a position of 4 \AA$ $ above the surface \cite{Odobesko2019, Zhussupbekov2021, Potorochin2022}. Changes in binding energy are calculated by comparing the value of the electric potential at the atomic centres as in Refs. \cite{Frolov2020, Yashina2008}. The work function was calculated by comparing the electric potential in the vacuum region $\phi_{\mathrm{vac}}$ of the slab with the Fermi level $E_f$ as in Ref. \cite{DeWaele2016}.

\subsection{B. Experimental Details}

All experiments were performed \textit{in-situ} on the same (110) terminated niobium single crystal. The crystal was annealed at \SI{850}{\celsius} under ultra-high-vacuum (UHV) conditions. During annealing the sample temperature was measured from a K-type thermocouple up to \SI{600}{\celsius}, with temperatures above \SI{600}{\celsius} estimated via an infrared optical pyrometer ($\epsilon$\,=\,0.25). The crystal was transferred between two UHV systems via a UHV suitcase with a base pressure of low $10^{-10}$ mbar. All STM images demonstrated were obtained with a commercial Createc slider-type STM in constant-current mode at 77\,K. The STM tips utilised were [001]-oriented single-crystalline W, which were electrochemically etched in NaOH. The bias was applied to the sample with respect to the tip. The UPS spectra were obtained with an excitation energy of He I (21.2 eV). XPS measurements were performed on an Omicron MultiProbe XPS system using monochromated Al K$_{\alpha}$ X-rays (XM 1000, 1486.7 eV) with an instrumental resolution of 0.6\,eV.

\section{III. Results}

\subsection{A. Crystal Structure}

To arrive at this model, we first consider a single isolated oxygen atom on top of the Niobium surface. Several absorption sites are examined, and our calculations confer with that of Tafen et al. \cite{Tafen2013} i.e. the oxygen atom sits in a 3-fold site on the surface. One might then expect that for a full monolayer of oxygen on top of the surface, the oxygen would simply fill the 3-fold sites as in Figure \ref{fig:fig1}(b). However, when a full monolayer of oxygen is added we see a reconstruction take place, with the surface Niobium atoms shifting to open up 4-fold sites on part of the surface. The relaxation of this structure is shown in Figure \ref{fig:fig1}(b-c), and is accompanied by an energy drop of 0.11 eV/atom. The only choice that we make when constructing this model is the size and shape of the supercell in which the structure is allowed to relax. We simply choose a supercell which matches the periodicity observed in the experimental STM images (see Figure \ref{fig:fig1}(a) and Figure \ref{fig:fig2}(a)) (i.e. 4 unit cells along bcc[-11-1]). Initially, we consider an ``infinite'' nanocrystal as shown in Figure \ref{fig:fig1}(c), where the pattern extends infinitely along the bcc[-111] direction and the disordered channels are not included. This is sufficient for the purposes of calculating the local properties of the nanocrystals such as STM images and STS curves. However, we do also perform calculations with finite nanocrystals (full unit cell including channels), which are necessary for calculating global electronic properties for comparison with area averaged techniques such as XPS. 

As shown in the top down crystal structure (Figure \ref{fig:fig1}(c, e)), the oxygen monolayer model displays a transition to the fcc (111) symmetry in the regions where the oxygen is 3-fold coordinated at the surface. In our calculations the epitaxial relationship closely resembles the Nishiyama-Wasserman (NW) orientation ((fcc)[1-10]$||$(bcc)[001]) observed in most experiments \cite{Delheusy2008, Zhussupbekov2020}. This is in agreement with the experimentally measured LEED pattern, shown in the inset of Figure \ref{fig:fig1}(a). Regarding the finite nature of the nanocrystals, previous STM studies have determined the average length of a nanocrystal to be between roughly 3.0 nm \cite{Razinkin2010, Surgers2001} to 3.5 nm \cite{Arfaoui2002}. Since this model still contains areas with fcc (111) symmetry, the previously proposed explanation for this behaviour based on rigid lattice theory \cite{Arfaoui2002, Razinkin2010} still holds. Along the nanocrystals (fcc [110]/bcc [-111]) the mismatch is roughly $\simeq$4-5\% between the lattices ($a_{\mathrm{fcc}}=2.98 \ \mathrm{\AA}$ and $\frac{\sqrt{3}}{2}a_{\mathrm{bcc}}=2.86 \ \mathrm{\AA}$), meaning after roughly $\simeq$10-12 unit cells the two lattices will be out of phase. This causes the finite rows with roughly 10-12 protrusions each, with length total length of ~ 3.0-3.5 nm.

\subsection{B. Local Electronic Structure}

\begin{figure*}
    \includegraphics[width=2\columnwidth]{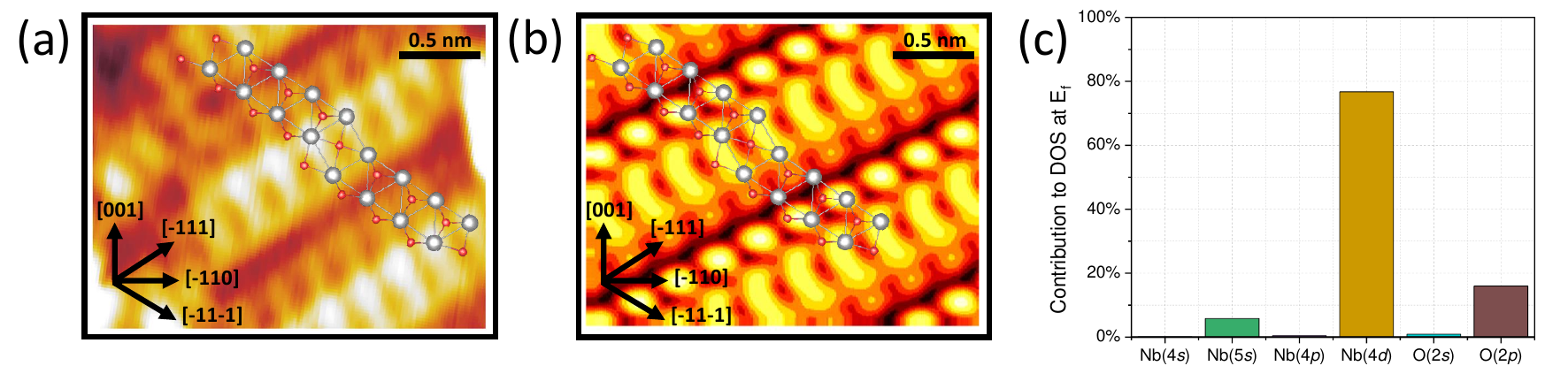}
    \caption{Comparison between experimental and simulated STM (a) Experimental STM image at bias 2 mV with suggested atomic structure superimposed. (b) Simulated STM image at a bias of 2 mV for the oxygen monolayer model with superimposed atomic structure. (c) Relative contribution of surface atom orbitals to the DOS at the Fermi level. 
}
\label{fig:fig2}
\end{figure*}

Figure \ref{fig:fig2}(a-b) shows an experimental STM image of the nanocrystal structure, as compared to the simulated STM image from the reconstructed oxygen monolayer model at the same bias (2 mV). In the simulated image we can see that the distinctive rows of wide protrusions are present, with the area in between these rows comparing favourably also. Since the ``monolayer of oxygen'' model does not include any extra adatoms or major height variation on the surface, it is interesting to consider the origins of the protrusions observed in STM. By considering the overlayed surface atomic structure on the simulated STM image in Figure \ref{fig:fig2}(b) we can see that the protrusions arise from the reconstructed 4-fold area of the surface. This can be understood looking at the calculated superimposed atomic structure. In these reconstructed 4-fold areas, the oxygen atoms sink deeper into the surface (see side view in Figure \ref{fig:fig1}(f)), leaving the Niobium atoms in this area more exposed to the vacuum. In the unreconstructed areas with 3-fold oxygen, the oxygen sits on top of the Niobium, passivating the Niobium orbitals that protrude from the surface, and reducing the LDOS at the Fermi level. Since almost all of the density near the Fermi level comes from the Niobium 4\textit{d} states (Figure \ref{fig:fig2}(c)), it is reasonable that the areas of the surface with the Niobium more exposed to the vacuum would show a higher LDOS around the Fermi level. The distinctive ``wide'' shape of the protrusions clearly arises from the 4-fold square nature of the reconstructed area.

\begin{figure*}[t!]
    \includegraphics[width=2\columnwidth]{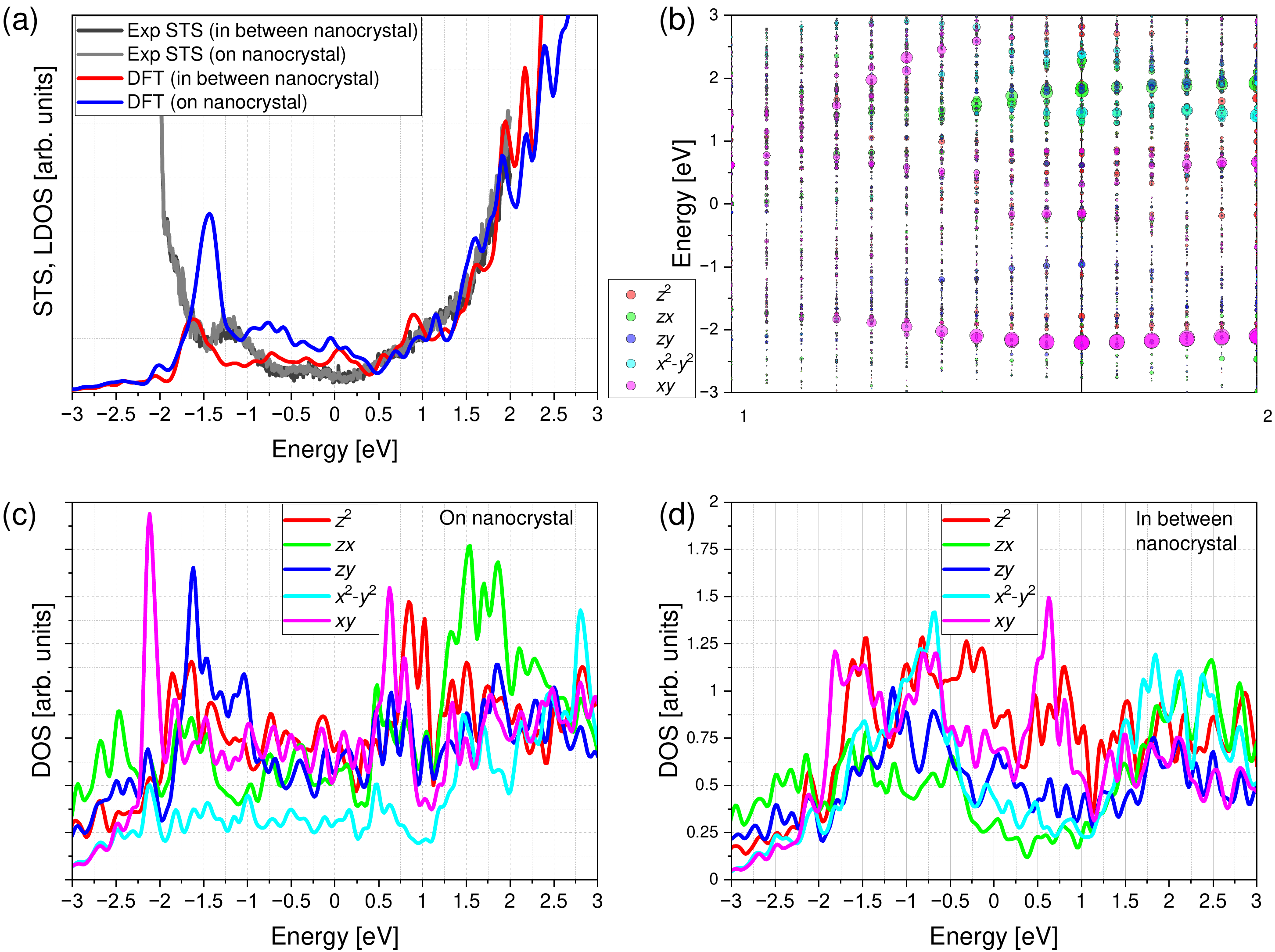}
    \caption{(a) Simulated LDOS at a distance 4 \AA\, above the surface, compared with measured experimental STS spectrum. (b) m$_l$ projected band structure for the surface nanocrystal atoms, (c-d) m$_l$ resolved DOS on and between the nanocrystal
}
\label{fig:fig3}
\end{figure*}

Turning to the STS, in Figure \ref{fig:fig3}(a) we can see good agreement between the experimentally measured STS and the LDOS as calculated in the vacuum above the slab for bias in the region -1 V to +2 V. In this region we observe minimal difference in the experimental STS on vs between the nanocrystal, and this is reflected also in the DFT simulation. The theoretical STS captures the relatively flat behaviour around the Fermi level from -0.5 to 0.5 eV, and the sharp increase above 0.5 eV. We see no evidence of resonances arising from surface states in the STS, such as was observed by Odobesko et al. \cite{Odobesko2019} for the clean Nb(110) surface. Our results for this oxygen saturated surface align with their findings that this feature is suppressed in oxygen rich areas, confirming that the resonance is unique to the clean Nb(110) surface. The agreement between the theoretical and measured STS breaks down only at large negative bias (below -1 V), where we see a resonance at -1.5 V in the theory, which is most pronounced on the nanocrystal. This lack of agreement below -1 V has been observed for the clean Nb(110) surface as well by Odobesko et al. \cite{Odobesko2019}, and is not unexpected as tip states tend to dominate the tunnelling matrix elements in this regime \cite{Odobesko2019, Zhussupbekov2020-1, Zhussupbekov2021-1}. 

In order to better understand the nature of the electronic states at this surface we plot the $m_l$ projected surface band structure (Figure \ref{fig:fig3}(b)). In the case of the nanocrystal structure, there are two equivalent N points (along the nanocrystal (N$_1$) vs across the nanocrystal (N$_2$)). We see clearly the absence of the \textit{z}$^2$ surface state observed by Odobesko et al. \cite{Odobesko2019} for the clean surface. Instead, the surface states below the Fermi level exist mostly around -2.2 eV and have a \textit{xy} orbital character. As expected, we do not see any resonance at -2.2 eV coming from these states in the theoretical STS (Figure \ref{fig:fig3}(a)) due to the orientation of the \textit{xy} orbitals, where the majority of electron density lies in the plane. We can understand the origin of the peak at -1.5 eV in the theoretical STS by looking at the $m_l$ projected density of states for the surface atoms on and between the nanocrystals (Figures \ref{fig:fig3}(c) and (d)) respectively. We see that around -1.5 eV, both contain a significant \textit{z}$^2$ component, but on the nanocrystal, there is a significant peak of \textit{zy} character, which is missing for the atoms between the nanocrystal. For these atoms between the nanocrystal, the in plane \textit{xy} orbitals play a more significant role. Therefore, we assign the resonance at -1.5 V observed in the theoretical STS to these \textit{zy} states. We note that this does not point to a \textit{zy} surface state, as can be seen from the band structure no such heavily localised state exists. The states responsible for this peak are bulk like, but with a high enough density to cause the peak in the calculated STS. The combination of the bulk like nature of these states, and the tip states dominating the tunnelling process in this regime, cause this feature to be unseen in the experimental STS. Overall, we see strong agreement between the DFT calculations and STM/STS experiments for this model.

\begin{figure*}[t!]
    \includegraphics[width=2\columnwidth]{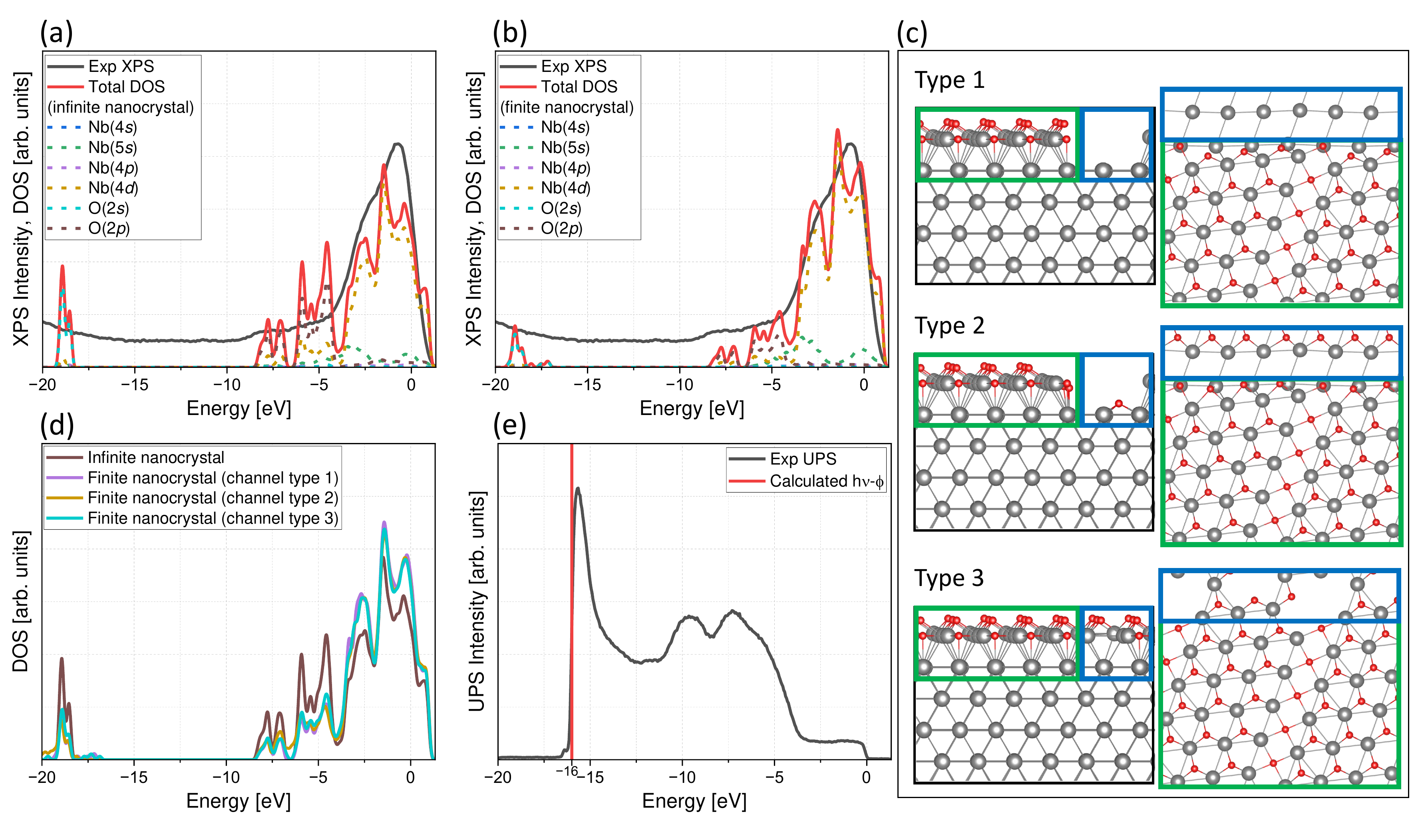}
    \caption{(a-b) Simulated total DOS over the whole slab with orbital decomposition, compared against measured valence band XPS for infinite and finite nanocrystals respectively. (c) Side and top views for the three different channel structures considered here, with the channel area highlighted by the blue rectangles and the nanocrystal area highlighted by the green rectangles. (d) Total DOS calculated for the three finite nanocrystal structures compared to infinite nanocrystal. (e) Experimental UPS spectrum compared with calculated value for the secondary electron cutoff $h\nu-\phi$.
}
\label{fig:fig4}
\end{figure*}

\subsection{C. Global Electronic Structure}

One of the most intriguing parts of this surface reconstruction is the measured core level binding energies. Despite the Nb/O ratio being close to 1\cite{Surgers2001,Arfaoui2002}, the Nb 3\textit{d} level shows an anomalous binding energy of 203.55 eV, a change in binding energy ($\Delta$B.E.) of 1.55 eV relative to the bulk \cite{Zhussupbekov2020, Razinkin2008}. This does not match up with any of the known bulk Niobium oxides. We calculate a change in binding energy for the surface Nb atoms of 1.43 eV relative to the bulk, in good agreement with the experimentally determined change. For the O 1\textit{s} level, two different chemical states \Oone and \Otwo are observed with energies 530.3 eV and 531.8 eV, $\Delta$B.E. of 1.5 eV with respect to one another \cite{Zhussupbekov2020, Razinkin2008}. The latter binding energy (531.8 eV) matches up exactly with the O 1\textit{s} binding energy in bulk NbO, while the former (530.3 eV) does not match with any bulk Niobium oxide. Within this oxygen monolayer model, the two chemical states are the 3-fold oxygen and the 4-fold oxygen in Figure \ref{fig:fig1}(c, f). For this change in binding energy, we calculate a value of 1.63 eV, again in good agreement with the experimental value. Because the exact ratio between the two chemical states of oxygen varies significantly between studies, we cannot compare this directly to the DFT calculations. However, in all studies there is always an asymmetry between the two chemical states, the \Oone chemical state is always more abundant than the \Otwo chemical state. This agrees with the oxygen monolayer model presented here, where the 3-fold oxygens correspond to \Oone, and the 4-fold oxygens correspond to \Otwo. Additionally, the \Otwo oxygen is slightly deeper into the surface (see side view in Figure \ref{fig:fig1}(f)). This accounts for the angle resolved XPS of Razinkin et al. \cite{Razinkin2008}, where they observe a higher \Oone/\Otwo ratio at glancing angle compared to normal incidence. Table \ref{tab:table1} shows the comparison between the experimentally fitted change in binding energy and the calculated values. Overall, there is strong agreement between XPS experiments and the oxygen monolayer model.

\begin{table}[b!]
\caption{\label{tab:table1}
Comparison of simulated and experimental core level binding energies and work function.
}
\begin{ruledtabular}
\begin{tabular}{l c @{\hspace{2\tabcolsep}} c}
\textrm{ }&
\textrm{Calculated}&
\multicolumn{1}{c}{\textrm{Experimental}}\\
\colrule
Nb 3\textit{d} $\Delta$B.E. [eV] & 1.43 \textrm{[this work]} & 1.55 \cite{Zhussupbekov2020,Razinkin2008}\\
O 1\textit{s} $\Delta$B.E. [eV] & 1.63 \textrm{[this work]} & 1.50 \cite{Zhussupbekov2020,Razinkin2008}\\
Work Function $\phi$ [eV] & 5.2 \textrm{[this work]} & 5.0 \textrm{[this work]}\\
$h\nu-\phi$ [eV] & 16.0 \textrm{[this work]} & 16.2 \textrm{[this work]}\\
\end{tabular}
\end{ruledtabular}
\end{table}

Looking now at the valence band XPS, we start by comparing the calculated DOS for the infinite nanocrystal model Figure \ref{fig:fig4}(a) to the experimentally measured XPS. We can see broadly good agreement between the two curves, however, the peaks in the vicinity of -5 eV give rise to a higher intensity than that observed in the experiment. In order to improve agreement between the valence band XPS and calculated DOS, we must account for the finite nature of the crystal by including the channels separating the nanocrystals (Figure \ref{fig:fig4}(b)). The finite nanocrystal including the channels drastically improves agreement between the simulated DOS and measured valence band XPS. The behaviour of the DOS around -5 eV is now fully captured, and agreement near the Fermi level remains strong. From the orbital resolved DOS in Figure \ref{fig:fig4}(a-b) we see that the main broad peak from 0 to -5 eV arises from mostly Nb 4\textit{d} states, whereas the shoulder around -5 to -7 eV comes mostly from the O 2\textit{p}. The deep feature around -18 eV is almost entirely arising from semicore O 2\textit{s} states.

Overall, it is not surprising that to obtain the correct DOS, the finite nature of the pattern must be included. While simulated STM images are highly local, and depend mostly on the local charge density on a surface atom, the total DOS will be highly sensitive to the total band structure across the entire Brillouin zone. When moving to the finite model the band structure will change dramatically, with the bcc[-111]/fcc[0-11] direction in the Brillouin zone (along the nanocrystals) shrinking by as much as a factor $\simeq$10. Therefore, a significant change in parts of the DOS is to be expected. It is also worth noting that since the channels separating the nanocrystals appear disordered in STM, simulating it using periodic DFT poses a challenge. For the purpose of ruling out the effect of the channel structure on the DOS we evaluate the 3 possible extremes of the channel structure. Figure \ref{fig:fig4}(c) shows the three structures of finite nanocrystals considered, with different atomic structures in the channels i.e., no oxygen/Niobium in the channel (i.e. a step edge) (Type 1), partial oxygen but no Niobium (Type 2), and partial oxygen/partial Niobium (Type 3). Figure \ref{fig:fig4}(d) shows the calculated DOS for these 3 structures compared to the infinite nanocrystal. We can see that while there is a clear difference between the infinite nanocrystal calculation (no channel) and the calculations including the channels, there is little difference between the calculated DOS for the different types of channel structures.

From our UPS experiment (Figure \ref{fig:fig4}(e)), the work function ($\phi$) of this surface can be obtained by comparing the incident photon energy to the secondary electron cutoff ($E_{\mathrm{sec}}=h\nu - \phi$). Using this method we obtain a work function of 5.0 eV (table \ref{tab:table1}). Comparing this result with the work function of Nb metal (4.3 eV) and Nb$_2$O$_5$ (5.2 eV) \cite{Tyagi2016}, we see the expected trend is upheld, with the work function increasing as more oxygen is added. Within DFT we can estimate the work function by comparing the classical electric potential at a point far from the surface slab with the Fermi level. From this method we obtain a calculated work function of 5.2 eV, showing decent agreement with the experimentally measured value. 

\section{IV. Conclusions}

From the above discussion, it seems clear that the oxygen monolayer surface is the best candidate for explaining the surface reconstruction. Due to the unusual structure of this surface, it is worth discussing how correct it is to call this surface structure a NbO structure. While the stoichiometry of the surface measured from XPS suggests the ratio of oxygen and Niobium to be very close to 1, the anomalous Nb 3\textit{d} core level binding energy suggests that the chemical environment of these atoms is not similar to that of bulk NbO. Our model confirms this, with none of the Niobium atoms being square planer coordinated (as they are in bulk NbO). Some papers have discussed this anomalous binding energy in terms of a sub-oxide Nb$_2$O \cite{Ma2003, Ma2004, Sebastian2006}. However, in these studies, the atoms assigned to that sub-oxide are buried underneath a Nb$_2$O$_5$ layer, and as previously mentioned the stoichiometry contradicts this sub-oxide interpretation. The crystallography also suggests that this layer is not simply a NbO monolayer. This surface therefore appears to be a novel 2D Niobium oxide layer, with no bulk counterpart. 

Having determined the precise local structure of this oxide layer, there is now a foundation in place upon which effect of the Niobium surface structure on device performance can be understood. This surface structure will be present on any Nb(110) facet, either buried underneath higher valence oxides or on its own. In the case of SRF cavities specifically, this surface structure may be the dominant surface oxide present in samples showing improved performance after vacuum annealing \cite{Ciovati2010}. Whatever the mechanism for the improvement in device performance may be, it is likely that this surface structure plays a significant role.

\section{Acknowledgements}
This work was supported by Irish Research Council (IRC) Laureate Award (IRCLA/2019/171), the Government of the Republic of Kazakhstan under the Bolashak program, the Russian Academy of Sciences through the state task of Institute of Solid State Physics RFBR Grant 19-29-03021 and Erasmus Plus mobility grants (2017-1-IE02-KA107-000538 \& 2018-1-IE02-KA107-000589). All calculations were performed on the Boyle cluster maintained by the Trinity Centre for High Performance Computing. This cluster was funded through grants from the European Research Council and Science Foundation Ireland. K.Z. and A.Z. would also like to acknowledge funding from IRC through GOIPD/2022/774 and GOIPD/2022/443 awards. This paper concerns work carried out from 2019 to 2021.

\section{Bibliography}
%

\newpage

\title{\huge{Supporting Information}\\[5mm] \Large {Unravelling the atomic and electronic structure of nanocrystals on superconducting Nb(110): Impact of the oxygen monolayer}}

\maketitle

\renewcommand\thefigure{SI\arabic{figure}} 
\setcounter{figure}{0} 

\onecolumngrid

\subsection{DFT simulation of literature models}
There have been two main models proposed in the literature, one from Razinkin \cite{Razinkin2010} and one from Arfouai \cite{Arfaoui2004}. Both of these models attribute the observed periodic arrays of protrusions to rows of Niobium atoms on top of a fcc bulk like NbO(111) plane with some arrangement of oxygen surrounding the structure. They differ in how they arrange these additional oxygen atoms. In the Arfaoui model a NbO(111) oxygen plane is stacked on top of the Nb(110) surface, followed by a NbO(111) Niobium plane, followed by another NbO(111) oxygen plane, and then finally the rows of Niobium atoms are placed on top in a row like arrangement. Bulk NbO takes on an unusual ``ordered defect'' structure \cite{Efimenko2017}, and these models include these ordered defects in their NbO layers. In the Razinkin model a NbO(111) Niobium plane is stacked on top of the Nb(110) surface, followed by a NbO(111) oxygen plane, followed by the rows of Niobium atoms, this time with additional oxygen atoms installed inside the nanocrystals. These additional oxygen atoms provide the second chemical state of oxygen O$_{\mathrm{II}}$. Again they include the ordered vacancy structure in the NbO(111) layers reminiscent of bulk NbO. The finite nature of the nanocrystals is explained in terms of the mismatch between NbO(111) and Nb(110), as explained previously.

Firstly, both of these models include the ordered vacancies present in bulk NbO in their surface layer (Figure SI1(a)), as well as Niobium adatoms, in their proposed surface layer. Razkinin et al. \cite{Razinkin2010} do mention the possibility of buckling of the nanocrystal in their model due to the vacancy, and our calculations indicate that when these ordered vacancies are present, the nanocrystal not only buckles but almost fully collapses, filling the vacancies (Figure SI1(c-d)). This can be understood for the following reason: in bulk NbO the ordered lattice of vacancies generate Nb$_6$ octohedra (Figure SI1(b)). The Nb \textit{d} orbitals can then overlap through these octahedra, leading to a release of delocalisation energy \cite{Hu2020,Schulz1993}. If this delocalisation energy is greater than the energy needed to create the ordered defects, this structure will be the most energetically favourable. It turns out that due to the dispersed nature of the Nb 4\textit{d} orbitals, this does in fact occur, so bulk NbO takes on the ordered vacancy structure as opposed to the rocksalt structure. However, at the surface, there is only a monolayer of NbO, oriented in the [111] direction, so there is no opportunity for Nb$_6$ octahedra to form. Thus, since there is no release of delocalisation energy, there is no reason to think that the ordered vacancy structure would be stable for a monolayer of NbO. This is corroborated by our DFT calculations. Therefore, we can say that models which include ordered vacancies cannot account for this surface reconstruction. Even if we modify these literature models to fill the vacancies, (which does lead to stable adatom nanocrystals) the agreement with experiment is still poor. Figures SI2 and SI3 show the structure and simulated STM/STS from these models (without vacancies). We can see these simulated STM images have poor agreement with the experimental images. The Niobium adatoms present in these models do not produce the distinctively shaped/wide protrusions observed in experiment. Hence, we can conclude that the observed protrusions in STM are not due to Niobium adatoms.

\newpage
\begin{figure}
    \includegraphics[width=0.6\columnwidth]{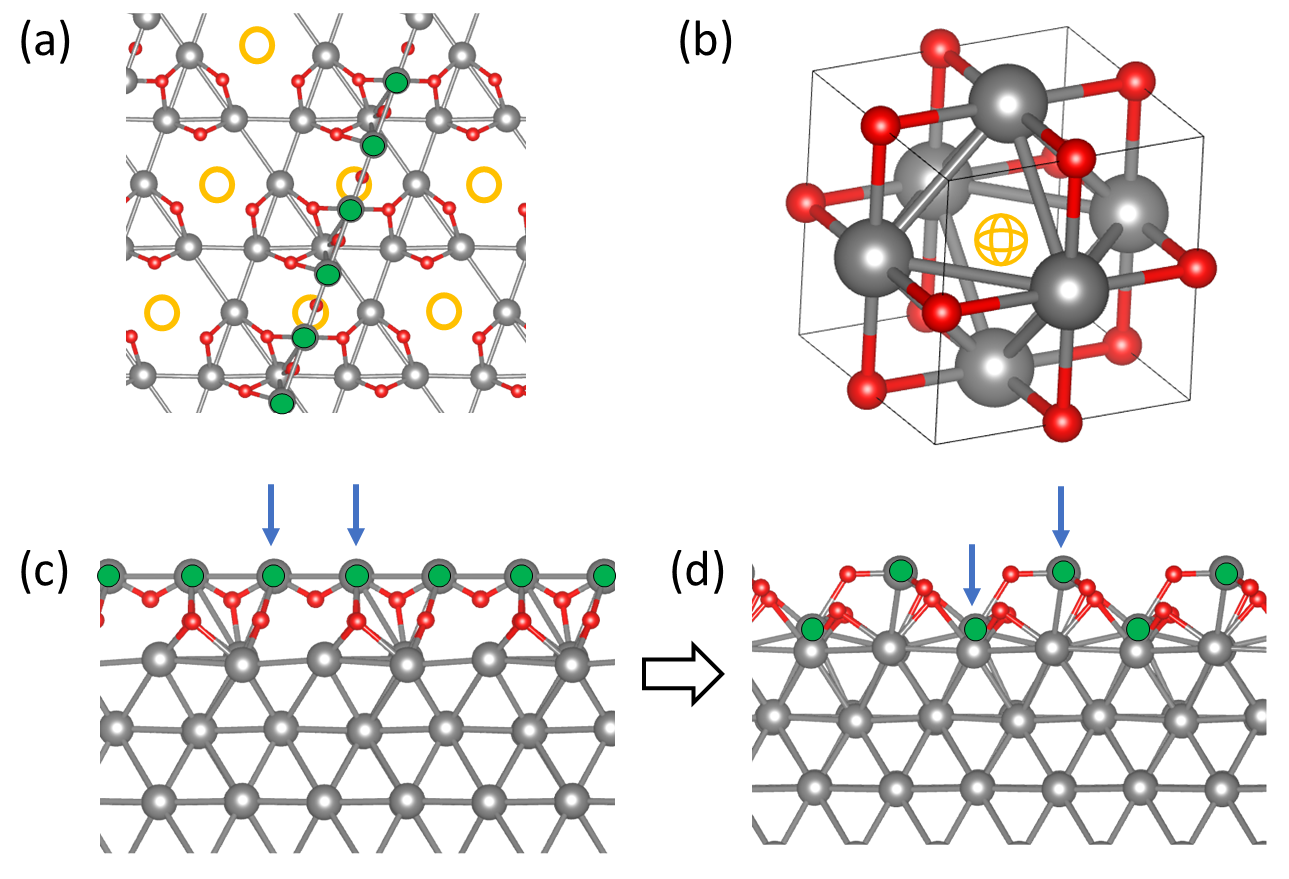}
    \caption{(a) Demonstrates a top down view of the Razinkin model, with nanocrystal atoms highlighted by the green circles, and Niobium vacancies highlighted by the gold circles. (b) Shows the bulk NbO crystal structure with the Nb vacancy at the centre of the Nb$_6$ octahedron highlighted by the yellow sphere. (c-d) show a side view of our DFT structural relaxation of the Razinkin model.}
\label{fig:afig2}
\end{figure}

\begin{figure}
    \includegraphics[width=0.7\columnwidth]{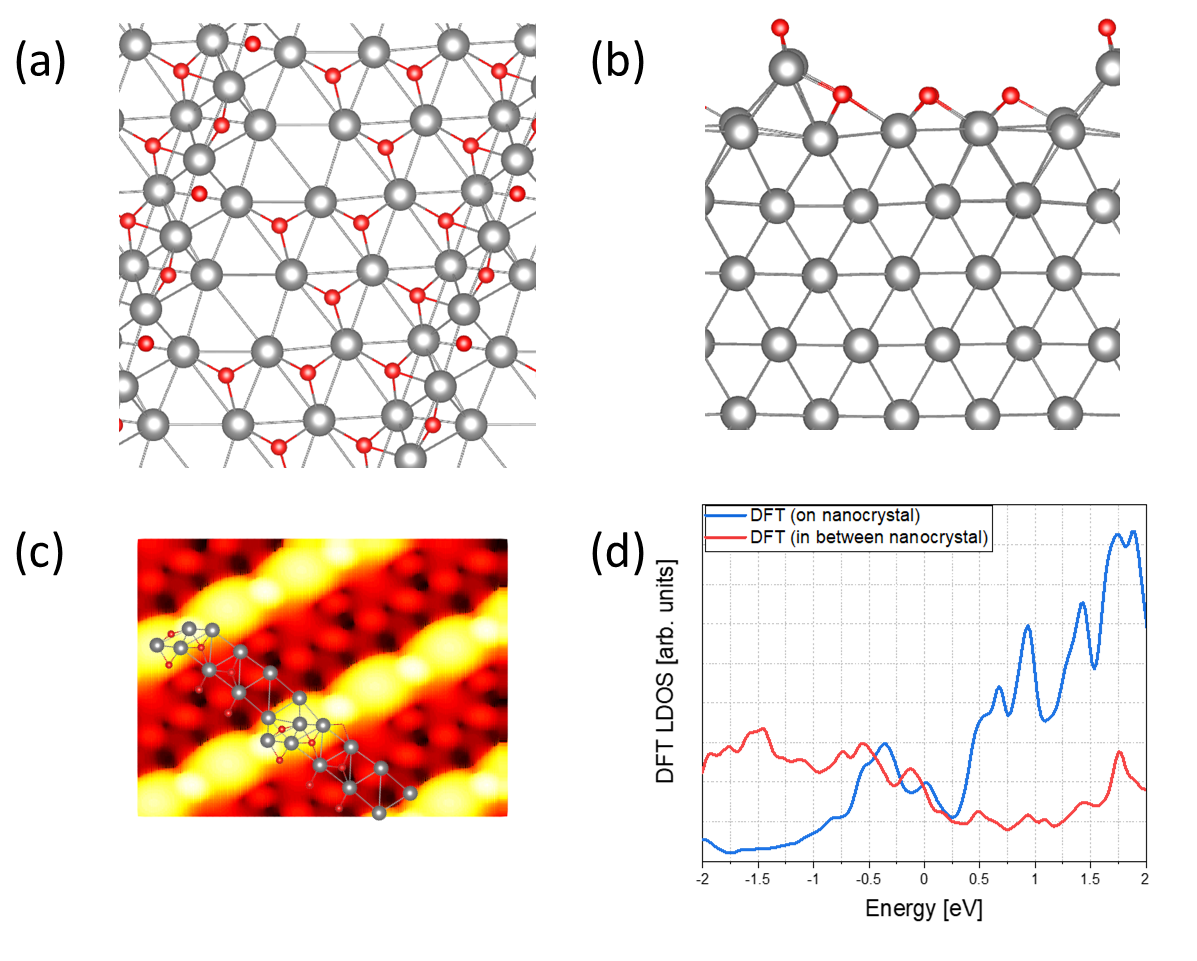}
    \caption{ Atomic structure and simulated STM/STS for Razinkin like model with vacancies filled.}
\label{fig:afig3}
\end{figure}

\begin{figure}
    \includegraphics[width=0.7\columnwidth]{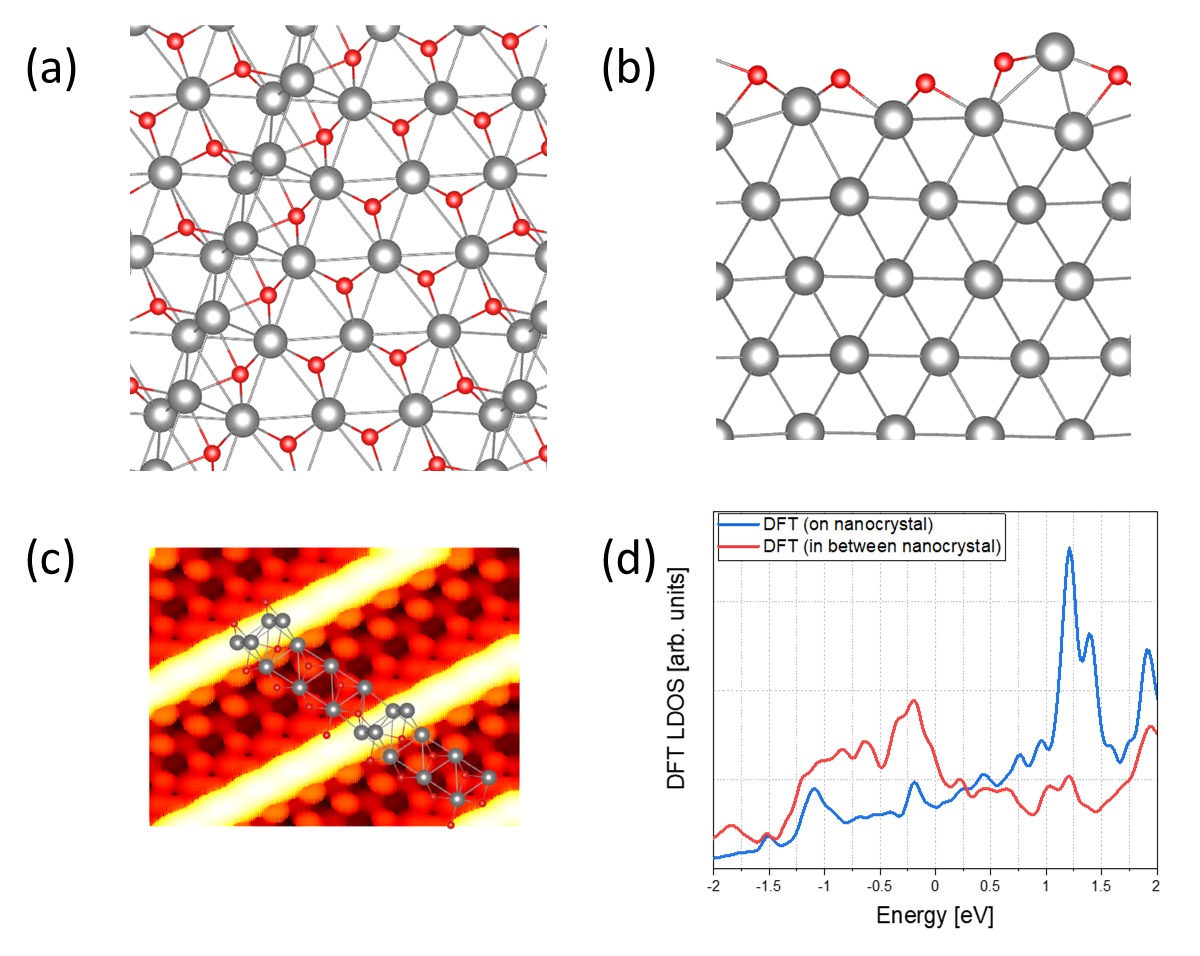}
    \caption{Structure and simulated STM/STS for an Arfoaui like model with vacancies filled and second oxygen layer removed.}
\label{fig:afig4}
\end{figure}

\end{document}